\documentclass{elsart}
\usepackage{epsfig}
\journal{Nucl. Instr. and Meth. A}

\begin{document}
\begin{frontmatter}
\title{BESII Detector Simulation}
\author[ihep]{M. Ablikim},
\author[ihep]{J.Z. Bai},
\author[pk]{Y. Ban},
\author[ihep]{J.G. Bian},
\author[ihep]{X. Cai},
\author[ihep]{J.F. Chang},
\author[ustc]{H.F. Chen},
\author[ihep]{H.S. Chen},
\author[ihep]{H.X. Chen},
\author[ihep]{J.C. Chen},
\author[ihep]{Jin Chen},
\author[hunan]{Jun Chen},
\author[ihep]{M.L. Chen},
\author[ihep]{Y.B. Chen},
\author[ccast]{S.P. Chi},
\author[ihep]{Y.P. Chu},
\author[ihep]{X.Z. Cui},
\author[ihep]{H.L. Dai},
\author[zhejiang]{Y.S. Dai},
\author[ihep]{Z.Y. Deng},
\author[ihep]{L.Y. Dong\thanksref{dongly}},
\author[tsinghua]{Q.F. Dong},
\author[ihep]{S.X. Du},
\author[ihep]{Z.Z. Du},
\author[ihep]{J. Fang},
\author[ccast]{S.S. Fang},
\author[ihep]{C.D. Fu},
\author[ihep]{H.Y. Fu},
\author[ihep]{C.S. Gao},
\author[tsinghua]{Y.N. Gao},
\author[ihep]{M.Y. Gong},
\author[ihep]{W.X. Gong},
\author[ihep]{S.D. Gu},
\author[ihep]{Y.N. Guo},
\author[ihep]{Y.Q. Guo},
\author[hawaii]{Z.J. Guo},
\author[hawaii]{F.A. Harris},
\author[ihep]{K.L. He},
\author[shandong]{M. He},
\author[ihep]{X. He},
\author[ihep]{Y.K. Heng},
\author[ihep]{H.M. Hu},
\author[ihep]{T. Hu},
\author[ihep]{G.S. Huang\thanksref{huanggs}},
\author[ihep]{X.P. Huang},
\author[shandong]{X.T. Huang},
\author[ihep]{X.B. Ji},
\author[ihep]{C.H. Jiang},
\author[ihep]{X.S. Jiang},
\author[ihep]{D.P. Jin},
\author[ihep]{S. Jin},
\author[ihep]{Y. Jin},
\author[ihep]{Yi Jin},
\author[ihep]{Y.F. Lai},
\author[ihep]{C.G. Li\thanksref{licg}},
\author[ihep]{F. Li},
\author[ccast]{G. Li},
\author[ihep]{H.H. Li},
\author[ihep]{J. Li},
\author[ihep]{J.C. Li},
\author[ihep]{Q.J. Li},
\author[ihep]{R.Y. Li},
\author[ihep]{S.M. Li},
\author[ihep]{W.D. Li},
\author[ihep]{W.G. Li},
\author[liaoning]{X.L. Li},
\author[nankai]{X.Q. Li},
\author[guangxi]{Y.L. Li},
\author[sichuan]{Y.F. Liang},
\author[huazhong]{H.B. Liao},
\author[ihep]{C.X. Liu},
\author[huazhong]{F. Liu},
\author[ustc]{Fang Liu},
\author[ihep]{H.H. Liu},
%\author[ihep]{H.M. Liu},
\author[ihep]{H.M. Liu\corauthref{cor}},
\corauth[cor]{Corresponding author}
\ead{liuhm@ihep.ac.cn}
\author[pk]{J. Liu},
\author[wuhan]{J.P. Liu},
\author[ihep]{R.G. Liu},
\author[ihep]{Z.A. Liu},
\author[ihep]{Z.X. Liu},
\author[ihep]{F. Lu},
\author[henann]{G.R. Lu},
\author[ustc]{H.J. Lu},
\author[ihep]{J.G. Lu},
\author[nanjing]{C.L. Luo},
% ...

\collab{(BES Collaboration)}

\address[ihep]{Institute of High Energy Physics, Beijing 100049, 
People's Republic of China}
\address[pk]{Peking University, Beijing 100871, People's Republic 
of China}
\address[ustc]{University of Science and Technology of China,
Hefei 230026, People's Republic of China}
\address[hunan]{Hunan University, Changsha 410082, People's Republic 
of China}
\address[ccast]{China Center for Advanced Science and Technology(CCAST),
Beijing 100080, People's Republic of China}
\address[zhejiang]{Zhejiang University, Hangzhou 310028, People's 
Republic of China}
\address[tsinghua]{Tsinghua University, Beijing 100084, People's 
Republic of China}
\address[hawaii]{University of Hawaii, Honolulu, Hawaii 96822, USA}
\address[shandong]{Shandong University, Jinan 250100, People's Republic 
of China}
\address[liaoning]{Liaoning University, Shenyang 110036, People's 
Republic of China}
\address[nankai]{Nankai University, Tianjin 300071, People's Republic 
of China}
\address[guangxi]{Guangxi University, Nanning 530004, People's Republic 
of China}
\address[sichuan]{Sichuan University, Chengdu 610064, People's Republic 
of China}
\address[huazhong]{Huazhong Normal University, Wuhan 430079, People's
Republic of China}
\address[wuhan]{Wuhan University, Wuhan 430072, People's Republic of 
China}
\address[henann]{Henan Normal University, Xinxiang 453002, People's 
Republic of China}
\address[nanjing]{Nanjing Normal University, Nanjing 210097, People's
Republic of China}
\address[shanghai]{Shanghai Jiaotong University, Shanghai 200030, 
People's Republic of China}
\address[guangxin]{Guangxi Normal University, Guilin 541004, People's
Republic of China}

\thanks[dongly]{Current address: Iowa State University, Ames, Iowa
50011-3160, USA.}
\thanks[huanggs]{Current address: Purdue University, West Lafayette,
Indiana 47907, USA.}
\thanks[licg]{Current address: 12175 Eastbourne Road, San Diego,
California 92128, USA.}
\thanks[shix]{Current address: Cornell University, Ithaca, New York 
14853, USA.}
\thanks[wangwf]{Current address: Laboratoire de l'Acc{\'e}l{\'e}ratear
Lin{\'e}aire, F-91898 Orsay, France.}
\thanks[xinb]{Current address: Purdue University, West Lafayette, 
Indiana 47907, USA.}
\thanks[zhaozg]{Current address: University of Michigan, Ann Arbor,
Michigan 48109, USA.}
\thanks[zhuyc]{Current address: DESY, D-22607, Hamburg, Germany.}

\begin{abstract}
A Monte Carlo program based on Geant3 has been developed for BESII 
detector simulation. The organization of the program is outlined, and 
the digitization procedure for simulating the response of various 
sub-detectors is described. Comparisons with data show that the 
performance of the program is generally satisfactory.
\end{abstract}


\end{frontmatter}

\section{Introduction}

To understand the intrinsic characteristics and performance of a
detector, for both detector design and physics analysis in 
experimental high energy particle physics, reliable Monte Carlo (MC) 
simulation is essential.

The Beijing Spectrometer (BES \cite{bes1,besch}) detector started to
take data in 1989 at the Beijing Electron Positron Collider (BEPC),
which operates in the center-of-mass energy range from 2 to 5 GeV. The
early simulation program was EGS \cite{egs} based and thus did not
generate hadronic interactions of the secondary particles in the
detector. Some effects, like fake photons produced by hadrons, were
not properly described. For the detector response, simple parametric
models for sub-detectors were used in the simulation. As the program
grew, the hard-coded structure made it difficult to maintain and make
improvements.

With the upgrade of the BES detector (BESII)~\cite{bes2} and BEPC in
1996, the development of a new Monte Carlo program, SIMBES, based on
the Geant3 package \cite{geant3}, was begun.  The decision was
motivated by many reasons.  First, larger data samples require better
detector simulation to reduce systematic uncertainties. Second, some
sub-systems of BESII were partly changed and some totally new, and it
was difficult to make the required software changes within the old
program structure. Meanwhile, the Geant framework had proven to be a
powerful tool for detector description and particle tracking in high
energy physics.

Geant3.21, the last version of Geant3, is used in SIMBES. Much effort
is made to model the signal generation, since the digitization
procedure is highly detector dependent. BESII data is used for tuning
parameters to describe the detector response.  Many physics channels
are compared with data to check the simulation, and the agreement is
reasonably good. In this paper, the general organization and features
of SIMBES are described in Section \ref{sect2}. Detailed simulations
of detector response, as well as comparisons with data, are presented
in Section \ref{sect3}.

\section{The BESII Simulation Program - SIMBES}\label{sect2}

\subsection{The BESII Detector}
BESII is a conventional solenoidal magnetic detector that is described
in detail in Ref. \cite{bes2}. A 12-layer vertex chamber (VC)
surrounding the beam pipe provides trigger and coordinate information.
A 40-layer main drift chamber (MDC), located radially outside the VC,
provides trajectory and energy loss ($\d E/\d x$) information for
charged tracks over 85\% of the total solid angle. The momentum
resolution is $\sigma_{p}/p=0.017\sqrt{1+p^2}$ ($p$ in GeV/c), and the
$\d E/\d x$ resolution for hadron tracks is $\sim 8\%$. An array of 48
scintillation counters surrounding the MDC measures the time-of-flight
(TOF) of charged tracks with a resolution of $\sim 200$ ps for
hadrons. Radially outside the TOF system is a 12 radiation length,
lead-gas barrel shower counter (SC). This measures the energies of
electrons and photons over ~80\% of the total solid angle with an
energy resolution of $\sigma_{E}/E=22\%/\sqrt{E}$ ($E$ in GeV).
Outside of the solenoidal coil, which provides a 0.4 Tesla magnetic
field over the tracking volume, is an iron flux return that is
instrumented with three double layers of counters which identify muons
with momentum greater than 0.5 GeV/c.

The geometrical and material descriptions of the detector are given in
Section \ref{sect3}. The experimentally measured field map is used in 
the simulation.

\subsection{Program Structure}
\begin{figure}[htb]
\begin{center}
\epsfig{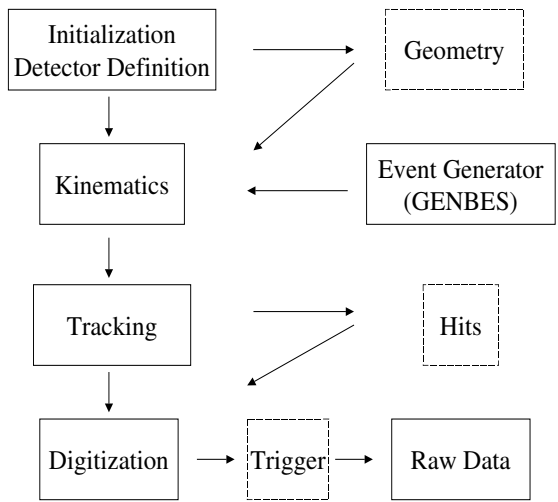}
\end{center}
\caption{Schematic structure of SIMBES, dashed boxes mean optional.}
\label{fig1}
\end{figure}

Built into the design of Geant is the separation of geometry, tracking, 
and digitization. SIMBES respects this feature as illustrated in 
Fig. \ref{fig1}. 
In the initialization stage, three kinds of data files are loaded: the
material and geometry configuration data used to define the detector,
the digitization constants to describe the detector response, and the
realization constants (such as dead channels and wire efficiency) to
simulate the real detector performance. Input values of these
parameters can be easily changed by experts or users using control
cards without relinking the program.

The BESII event generator package (GENBES) is a stand-alone program
that can be executed from within SIMBES to provide kinematic quantities 
for primary events. The generators for radiative QED processes
\cite{berends}, selected decays of $J/\psi$, $\psi(2S)$, and charmed
meson pair production \cite{mark3} were taken from other
experiments. The BES developed generators, LUND-CRM for inclusive
decays of $J/\psi$ and $\psi(2S)$ \cite{chenjc} and PPGEN for some
exclusive decays of the $\psi(2S)$, as well as some standard
generators like JETSET, are also included in GENBES.

At the end of each stage of processing (see Fig. \ref{fig1}), the 
intermediate output (the geometry or hit information in the Geant data 
structure) can be saved to a disk file, and used as input for later 
SIMBES execution. SIMBES code structure is hierarchical 
and modularized. Therefore, each sub-detector stands alone and can be 
turned on or off by the user. The BESII 
trigger logic is also implemented in a single module; the user has an 
option to switch it on or off when the digitization stage is finished. 
SIMBES is usually run in batch mode to generate raw data and ``MC-truth'' 
data. The interactive mode is used for graphic event 
display.

\subsection{Tracking and Hits}
The tracking of particles through the BESII detector is essentially
done according to the Geant standard. Tracking cutoffs and parameters
are carefully tuned for each sub-detector to provide a compromise
between simulation precision and computing speed. To correctly
simulate electromagnetic shower development in the shower counter,
the kinetic energy cutoffs of electrons and photons have been set to
Geant's limit (10 KeV).

When a charged particle traverses a sensitive volume (a cell or 
readout channel), the hit information, such as the particle position, 
energy deposit, etc., are saved for the digitization procedure which 
is discussed in the next section.

\subsection{Hadronic Interaction}
For hadronic interactions, several models have been tried and compared 
with BESII data; each has advantages for specific channels. 
Fluka \cite{Fluka} and Gcalor \cite{gcalor} are the two hadronic 
models that users can choose in SIMBES. 

\begin{figure}[htb]
\begin{center}
\epsfig{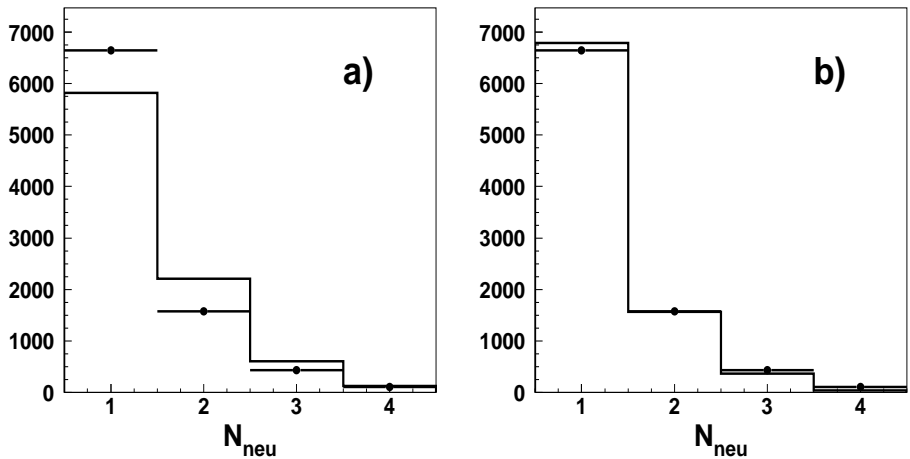}
\end{center}
\caption{Number of neutral tracks in the shower counter for
$\psi(2S)\rightarrow\pi^+\pi^-J/\psi\rightarrow\pi^+\pi^-\mu^+\mu^-$
decays. Histograms are for MC data with (a) Fluka or (b) Gcalor, and 
points with error bars are for data.}
\label{fig2}
\end{figure}

Extra photons produced by hadronic interactions in the shower counter
are studied by a very clean channel
$\psi(2S)\rightarrow\pi^+\pi^-J/\psi\rightarrow\pi^+\pi^-\mu^+\mu^-$
(only 4 charged tracks in the final state). The neutral tracks
observed come mainly from pion hadronic interactions in the SC
material. Fig. \ref{fig2}
shows comparisons between data and MC data for different 
hadronic packages. Gcalor reproduces the number of 
neutral tracks in the SC better than Fluka.

\begin{figure}[htb]
\begin{center}
\epsfig{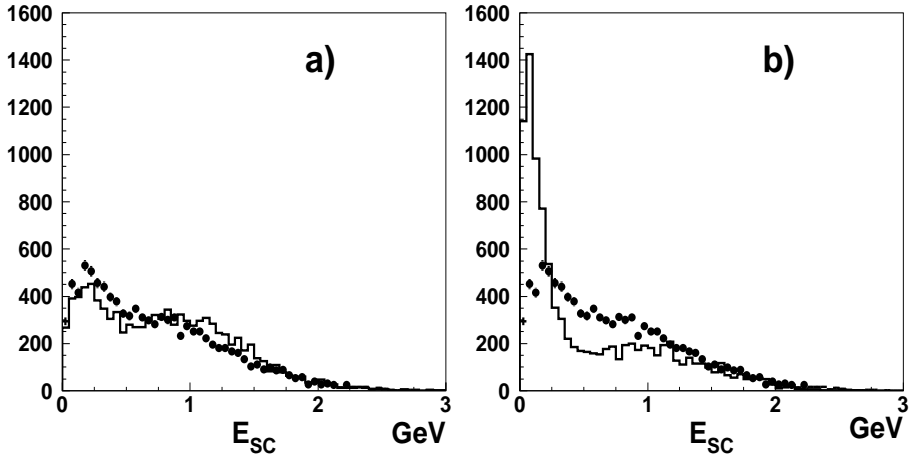}
\end{center}
\caption{Energy deposited by $\bar{p}$ in the shower counter for
$J/\psi\rightarrow\pi^+ \pi^- p \bar{p}$ decays.
Histograms are for MC data with (a) Fluka or (b) Gcalor, and points with
error bars are for data.}
\label{fig3}
\end{figure}

In Fluka, inelastic cross sections for anti-nucleon-nucleus
annihilation are set to zero if the anti-nucleon ($\bar{n}$ or
$\bar{p}$) kinetic energy is below 50 MeV (corresponding to a momentum
of 310 MeV/c); this is, of course, not satisfactory for the BEPC
energy domain.  Therefore, the parametrization given in \cite{flukac}
is used instead to approximate the cross-sections for anti-nucleons at low
momentum. For other particles and for elastic cross-sections, standard
Fluka is used.  Fig. \ref{fig3}
shows the energy deposition in the SC for low momentum (less than 0.8 
GeV/c) anti-protons for $J/\psi\rightarrow\pi^+ \pi^- p \bar{p}$ 
decays. Fluka's simulation is better than Gcalor's.

\section{Simulation of Detector Response}\label{sect3}

\subsection{Vertex of event and Vertex Chamber}
The location of the primary event vertex is determined by the
interaction point (IP) of the beam bunches, which varies with each beam
collision. The IP location and beam size distributions are obtained from
experimental data for different running periods, and they are used to 
sample the primary vertex in MC event generation. A $T_0$ fluctuation 
caused by the extended beam bunch size is taken into account in the TOF 
simulation (see Section \ref{sect3.3}).

\begin{figure}[htb]
\begin{center}
\epsfig{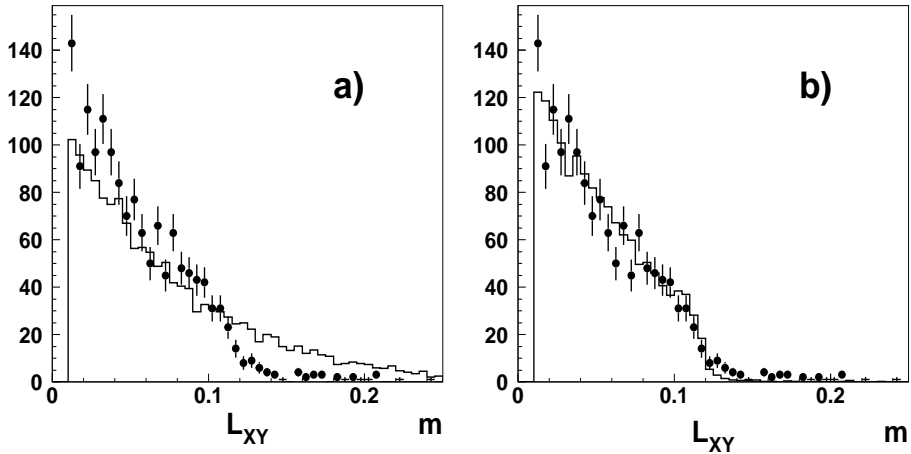}
\end{center}
\caption{Comparison of the $K_S^0$ (from $J/\psi\rightarrow K_SK_L$ 
decays) decay length distributions between data and MC data. Histograms 
are for MC, and points with errors for data (a) without the trigger and 
(b) with the trigger.}
\label{fig4}
\end{figure}

The vertex chamber (VC), which is built around the beryllium
beam-pipe, is composed of 12 
layers of straw-tubes. Its hit (time) information is used for 
the trigger. When a 
charged particle traverses a straw-tube, the minimum distance between 
the track and the wire is taken as the drift distance and smeared
according to 
the wire space resolution. The drift time of the ionization electrons 
is obtained by dividing the drift distance by an effective drift 
velocity. The wire efficiency and the wire sag due to gravity 
are also taken into consideration in this calculation. The VC simulation 
is checked using the decay $J/\psi\rightarrow K_{S}K_L$ \cite{kskl},  
which may lose events in the VC trigger because of the long decay
length of the $K_{S}$ particles. The agreement between data and MC
simulation is good when 
the VC trigger is applied (Fig. \ref{fig4}).

\subsection{Main Drift Chamber}
The main drift chamber, which is used to determine trajectories and
measure energy losses of charged particles, is the main tracking
detector of BESII. It is cylindrical with 10 super layers with 4 sense 
wires (axial or stereo) in each superlayer. The axial layer is
described by Geant's TUBE volume. For stereo layers, the shape of the
hyperbolic tube (HYPE) in Geant cannot reproduce the geometry.
Instead, the subtraction technique is used on two HYPEs for the
stereo-layer description. In the simulation, the wires (sense, field
and guard) are not put into SIMBES individually, but an equivalent 
material inside the MDC, taken as a mixture of gas and wires, is used.

\begin{figure}[htb]
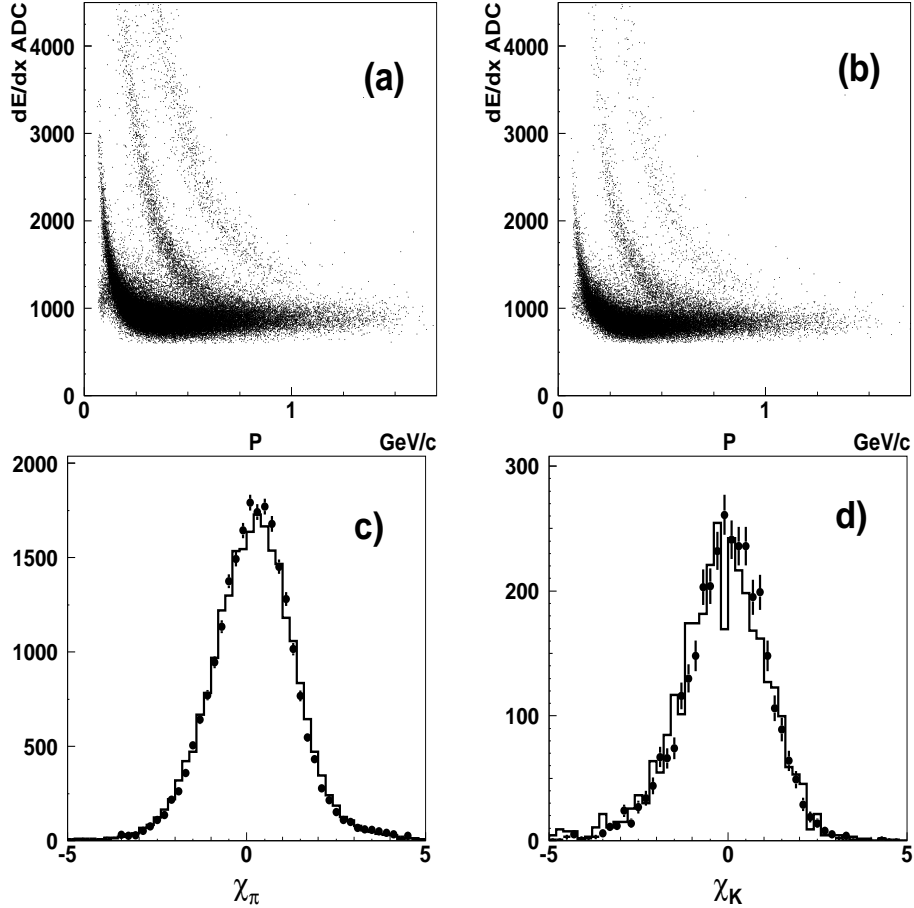

\begin{center}
\epsfig{file=fig5-1.epsi,height=6.0cm,width=12.0cm}
\epsfig{file=fig5-2.epsi,height=6.0cm,width=12.0cm}
\end{center}
\caption{Top: $\d E/\d x$ (pulse-height) versus momentum for (a) MC
and (b) data. Bottom: The
$\chi_i$ $((\d E/\d x)_{measured}-(\d E/\d x)_{expected})/\sigma)$
distributions for (c) $\pi$  (d) $K$ particles from inclusive $J/\psi$
decays. Histograms are for MC data, and points with error bars are for 
data.}
\label{fig5}
\end{figure}

The energy loss calculated by Geant for a thin layer is not used in
simulating the energy loss by charged tracks in the MDC. Instead, the
energy loss (Q) at a sense wire is simulated according to the
distribution obtained from real data, smeared with an
experimentally determined Landau distribution.  Fig. \ref{fig5}
shows $\d E/\d x$ distributions for different particles, where $\d
E/\d x$ is the truncated mean of the energy loss for each track.

\begin{figure}[htb]
\begin{center}
\epsfig{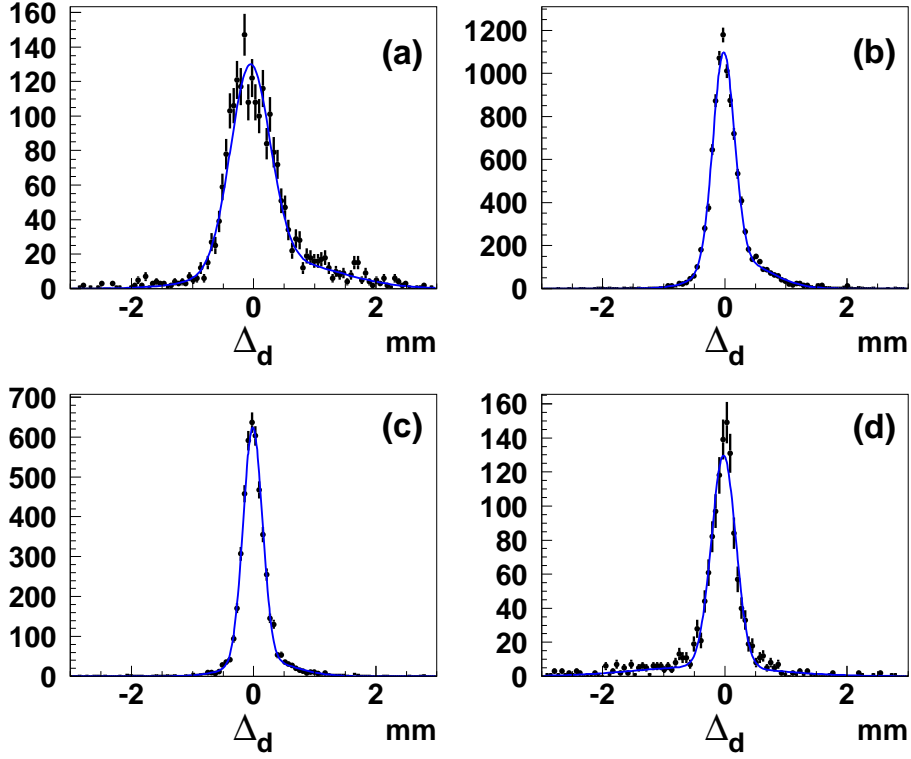}
\end{center}
\caption{Distributions of drift distance residuals $\Delta_d$ for
different pulse heights, Q, corresponding to ADC
counts of (a) 100-500, (b) 900-1100, (c) 1900-2100, and (d) 3700-4000.
Points with error bars are from Bhabha events at the $J/\psi$, and 
curves are fits with combined distributions of two Gaussians.}
\label{fig6}
\end{figure}

Using wire hit information, the minimum distance between a track and 
each wire is calculated including the effects of the Lorentz angle
($\alpha_L=26^o$) and gravitational sagging of the wires. The drift 
distance of the ionization is taken as the minimum distance and is
smeared by the wire space resolution ($\sigma _w$). It is observed
from data that the residual of the drift distance is dependent on Q, 
and even for a given Q, it cannot be simply fitted with a single 
Gaussian (see Fig. \ref{fig6}). 
Therefore, the drift distance is smeared 
according to a combined distribution of two Gaussians (fit from data) 
with the amount of smearing determined by the Q value and tracking 
layer. The wire efficiency is included in the $\sigma _w$ simulation. 
When there are multiple hits in a cell, the time signal is taken as the 
minimum drift time, while the pulse height is determined from the sum 
of energy losses.

\begin{figure}[htb]
\begin{center}
\epsfig{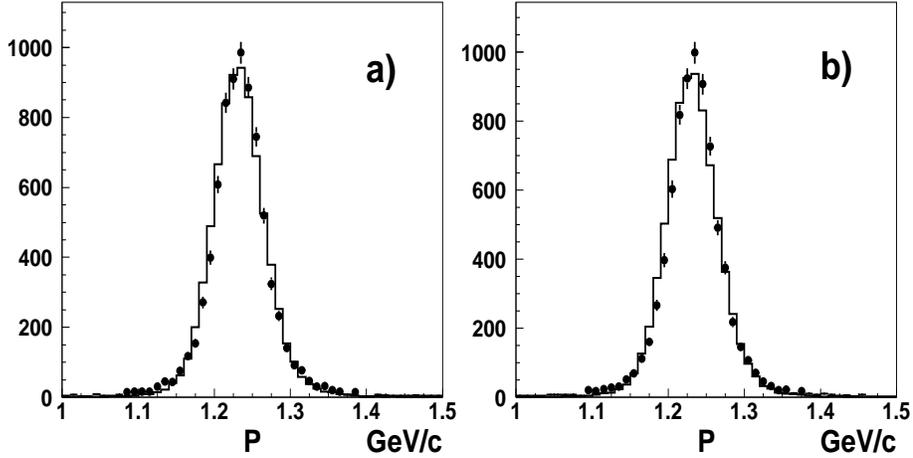}
\end{center}
\caption{Momentum distributions for (a) protons  and (b) anti-protons
from $J/\psi\rightarrow p \bar{p}$ decay. Histograms are for MC, and
points with error bars are for data.}
\label{fig7}
\end{figure}

\begin{figure}[htb]
\begin{center}
\epsfig{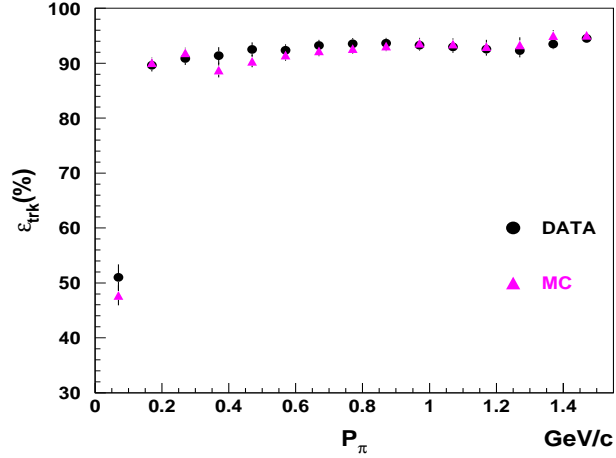}
\end{center}
\caption{$\pi$ tracking efficiency in the MDC versus momentum.
The first three points are from $J/\psi\rightarrow\Lambda
\bar{\Lambda}\rightarrow\pi^+ \pi^- p \bar{p}$,  and the others are from
$J/\psi\rightarrow\pi^+ \pi^- \pi^0$. }
\label{fig8}
\end{figure}

Fig. \ref{fig7} 
shows momentum distributions (after track reconstruction) for data 
and MC data. The comparison of momentum resolutions for data and Monte 
Carlo is shown in Table \ref{tab1}.
\begin{table}[htb]
\begin{center}
\caption{Comparison of momentum resolutions for data and Monte Carlo data.}
\label{tab1}
\begin{tabular}{c|c|c} \hline \hline
Channel & $\sigma_{MC}$ (MeV/c) & $\sigma_{DT}$ (MeV/c)\\ \hline
$e^+e^-\rightarrow\mu^+\mu^-$ at $J/\psi$ & 47.01 $\pm$ 0.06 & 46.70
$\pm$ 0.08\\ \hline
$J/\psi\rightarrow p\bar{p}$ & 33.22 $\pm$ 0.28 & 34.18 $\pm$ 0.12\\
\hline
\hline
\end{tabular}
\end{center}
\end{table}
The pion tracking 
efficiency (see Fig. \ref{fig8}) for simulated data 
is also found to be consistent with data.  In general, an error of 2\%
per track is taken as the systematic error from the tracking efficiency.
Alternative algorithms for the MDC wire resolution simulation have
also been tested, and their performances are slightly worse; however,
they can be used to determine the systematic uncertainties from the
MDC wire resolution simulation.

\subsection{Time of Flight System}\label{sect3.3}
The TOF system provides precise time measurements for particle 
identification and the trigger. It consists of two parts, the barrel 
(BTOF) and the end-caps (ETOF). The BTOF, located just outside of the 
MDC, employs 48 scintillation counters with photo-multiplier tubes 
(PMT) at each end. Each ETOF contains 24 scintillators with PMTs at one 
end. In SIMBES, the TOF system is described simply by trapezoidal 
scintillator bars with small gaps between adjacent ones.

After the track passes through the scintillator, the TOF hit information
contains the particle's position, $Z$, at the entry point, its energy 
deposition, $E_{loss}$, in the scintillator, and the time of flight, 
$T_f$, from the IP. $E_{loss}$ is determined using energy loss with 
$\delta$-ray generation and applying Birk's law for saturation, while 
for $T_f$, the time spread due to the sizable beam bunch is taken into 
account. A parameterized model is used to simulate the ADC and TDC 
outputs; it assumes the light produced at the entry point propagates with
attenuation to the two ends. The amplitude and arrival time are
obtained according to
\begin{equation}
Q = f\times E_{loss}\times\exp(Z/L_{atte}), T = T_f + Z/v_c + T_c(Q,Z),
\end{equation}
where f is a conversion constant, $L_{atte}$ the attenuation length 
(about 3m), $v_c$ the effective light velocity in scintillator, the 
last term in $T$ is a correction for $Q$ and $Z$ which is obtained from 
the calibration.  The time walk effect is included in $T_c$.

ADC and TDC counts are generated by smearing $Q$ and $T$ according to
their resolutions, and a $Q$ threshold is applied to the signal. The
resolutions for each counter are obtained using Bhabha events; in the
smearing, the time resolution dependence with amplitude and hit
position is taken into account. For multiple hits in a single counter,
taking the time walk effect as a good approximation, only the hits
with $Q$ over threshold contribute to the TDC count, while all pulse
heights are summed to give the ADC count.

\begin{figure}[htb]
\begin{center}
\epsfig{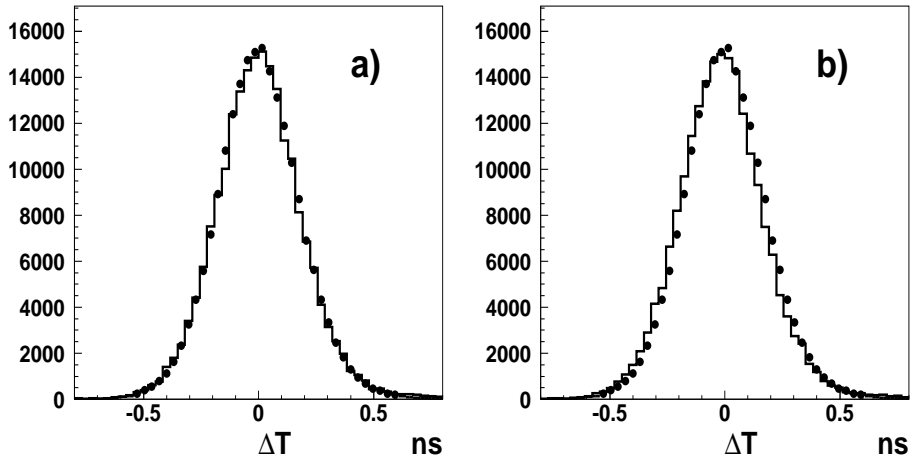}
\end{center}
\caption{$\Delta T$ $(T_{measured}-T_{expected})$ distributions
for (a) Bhabha and (b) dimuon events at the $J/\psi$ energy. Histograms 
are for MC data, and points with error bars are for data.}
\label{fig9}
\end{figure}

\begin{figure}[htb]
\begin{center}
\epsfig{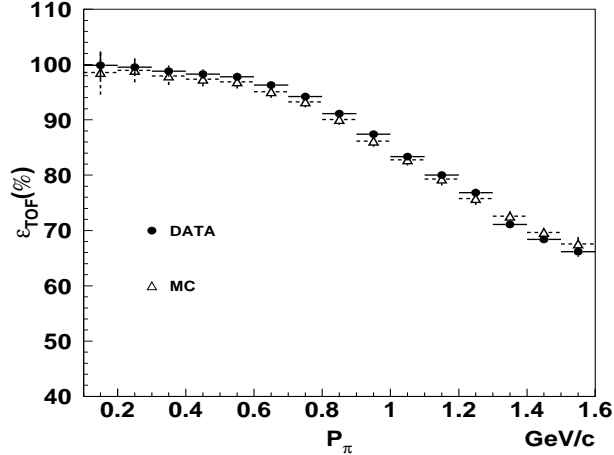}
\end{center}
\caption{Pion identification efficiency as a function of momentum
for the TOF from
  $J/\psi\rightarrow\pi^+ \pi^-\pi^0$).}
\label{fig10}
\end{figure}

Fig. \ref{fig9} 
shows the $\Delta T$ distributions for Bhabha and dimuon
events, and the pion particle identification (PID) efficiency for the BTOF 
as a function of momentum is shown in Fig. \ref{fig10}. 
The simulation is in a good agreement with data, and an uncertainty of
1\% per track is usually taken as the systematic error for PID with the
TOF.

\subsection{Shower Counter}
The shower counter consists of two parts, the barrel (BSC) and the
end-caps (ESC). In this section, BSC is taken as an example to 
demonstrate the simulation procedure. The cylindrical BSC consists of 
24 layers of gas tubes interleaved with 23 layers of lead absorbers. 
There are 560 cells (tubes made of Aluminum) in each gas layer. In 
order to reduce the number of electronic readout channels, layers 
with the same $\phi$ angle are grouped into six readout layers in the 
$r$ direction. To save computation time, the absorber layer is modeled 
using a mixture of Al-Pb-Al sandwiches as its material, while the 
sensitive layer (cell) is filled with gas working in self-quenching 
streamer (SQS) mode. The support ribs and some insensitive regions 
around them are properly described in SIMBES.  

When a charged particle hits a  gas tube,  the energy deposited (total 
charge) is sampled according to the SQS spectrum which can be 
parameterized by a Landau distribution. Corrections are made to this 
charge for the angle dependence of the track and for different layers. 
The pulse heights at the two ends of the tube are derived by charge 
division, and the final ADC outputs are obtained by smearing the pulse 
heights with a Gaussian which accounts for the contribution from the
electronics. In the case of multiple hits in a tube, if two hits are 
close enough, they are merged as one hit.

\begin{figure}[htb]
\begin{center}
\epsfig{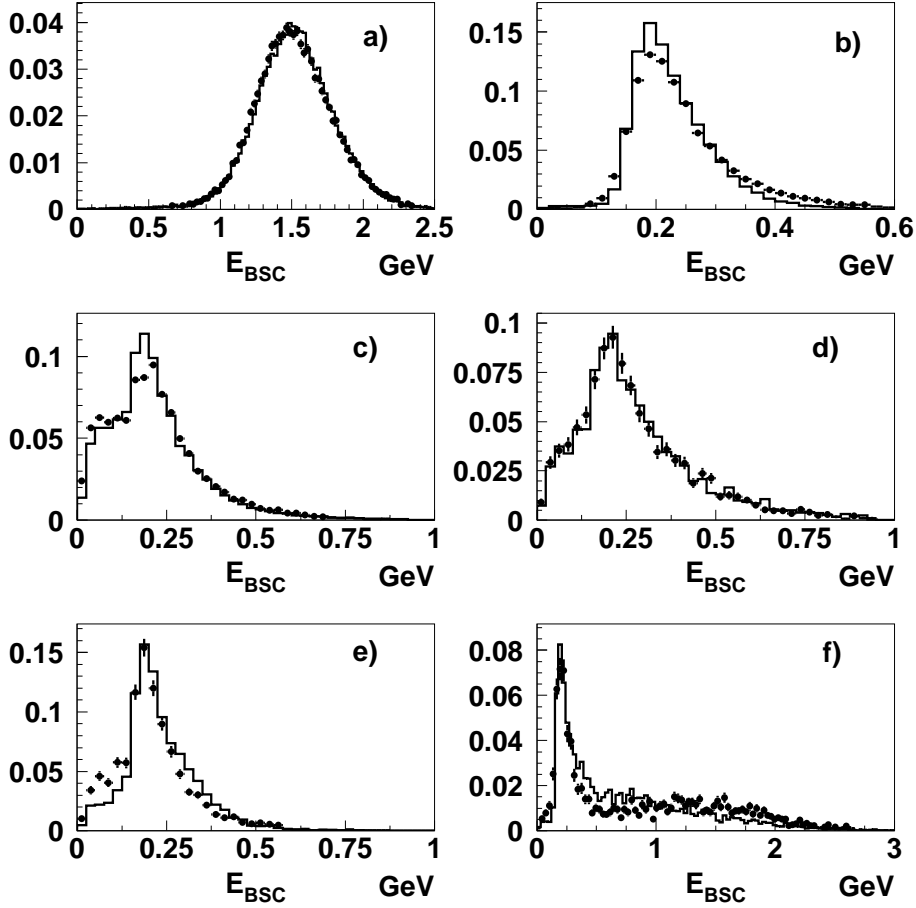}
\end{center}
\caption{Particle energy deposited in the BSC, (a) and (b) for $e$ and
$\mu$ from Bhabha and dimuon events at the $J/\psi$, (c) and (d) for
$\pi$ and $K$ from inclusive $J/\psi$ decays, and (e) and (f) for $p$ and
$\bar{p}$ from $J/\psi\rightarrow p\bar{p}$. Histograms are for
MC data,  and points with error bars are for data.}
\label{fig11}
\end{figure}

\begin{figure}[htb]
\begin{center}
\epsfig{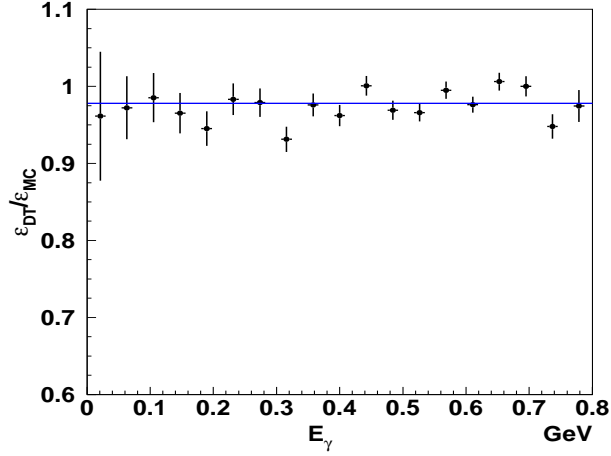}
\end{center}
\caption{The ratio of photon detection efficiency between data
and MC ($\epsilon_{DT}/\epsilon_{MC}$) versus photon
energy. $\epsilon_{DT}$ and $\epsilon_{MC}$ are calculated using
$J/\psi\rightarrow\pi^+ \pi^-\pi^0$ decays. The solid line is a fit
to the points.}
\label{fig12}
\end{figure}

The simulated hit profiles (number of hits in different layers) of an 
electromagnetic shower are in close agreement with data. 
Fig. \ref{fig11} 
shows the energy deposition (using Gcalor) in the BSC 
for different particles after the energy calibration with Bhabha 
events. Fig. \ref{fig12} \cite{bsc} 
shows the ratio of photon 
detection efficiency in the BSC for data and MC data. The comparison 
indicates reasonable agreement, and 2\% per photon is taken as the
systematic error for the photon detection efficiency.

\subsection{Muon Counter}
The muon identifier is the outermost component of BES and is composed 
of three layers of iron absorber and three layers of proportional 
chambers. There are 189 chambers distributed in an octagonal structure, 
each chamber consists of 8 proportional gas tubes which are arranged 
into two sub-layers. The complex geometrical structure and magnetic 
field \cite{field} of the muon system are well described in SIMBES.

\begin{figure}[htb]
\begin{center}
\epsfig{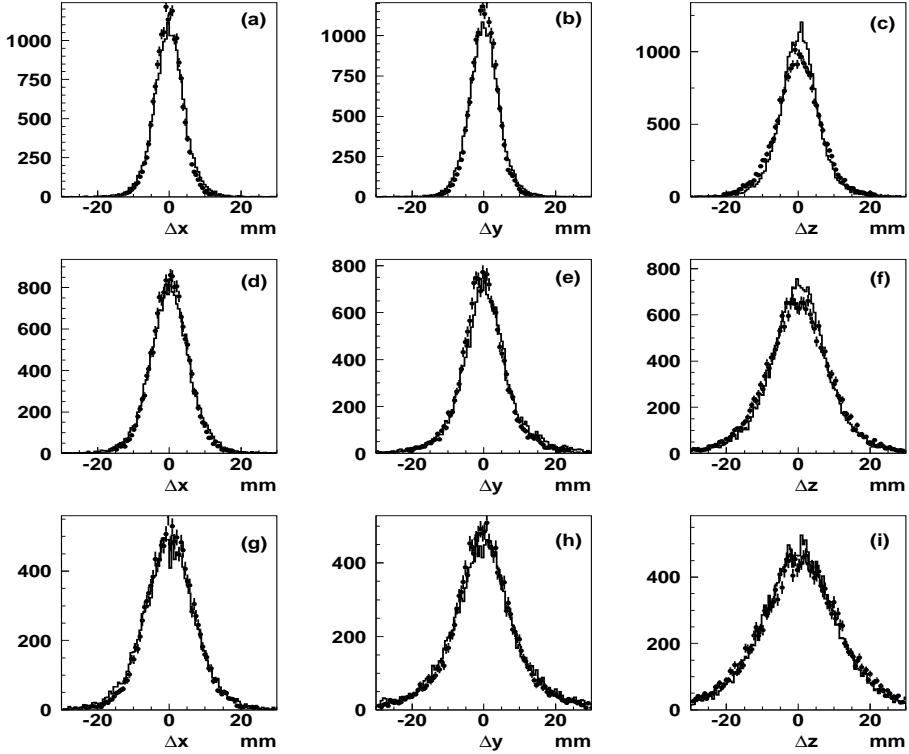}
\end{center}
\caption{ Hit position residuals for three Muon layers; the top, middle, 
and bottom plots correspond to layers 1, 2, and 3, respectively. Muons 
are from $e^+e^-\rightarrow\mu^+\mu^-$ at $J/\psi$ events. Histograms 
are for MC data, and points with error bars are for data.}
\label{fig13}
\end{figure}

In the simulation, the hit position (Z) along the tube is smeared by
the resolution (obtained from data), the ADC signals at the two ends
are obtained from charge division of the energy deposit from Geant,
and the wire efficiency from data is also included. Fig. \ref{fig13}
shows the comparison of the hit positions in the muon chamber for data
and MC data. The agreement is satisfactory.

\section{Summary}
A Monte Carlo program SIMBES has been developed for BESII 
simulation. Special efforts are made to accurately simulate the response 
of the detector.  For each sub-detector, some key physics quantities are
compared with BESII data to test the simulation quality. It is shown
from many checks that the overall performance of the current SIMBES is
satisfactory and that the results from simulation are generally 
consistent with data. On the other hand, there is still room for further
improvement. For example, the digitization of MDC should
be regarded as an `effective' simulation where many physics mechanisms
are hidden, and noise in the MDC is not considered at present.

The new Monte Carlo package (SIMBES) has significant impact on 
physics results. As an example, the branching ratio of 
$J/\psi\rightarrow\pi^+ \pi^-\pi^0$ measured by BES is 
$(2.10\pm0.12)\%$ \cite{phys}, which is significantly higher than the 
PDG value of $(1.50\pm0.20)\%$ \cite{pdg}. In this measurement, it is 
found that the efficiency using a two-Gaussian resolution function 
in the MDC wire simulation is about 20\% lower than that using a single 
Gaussian function, and that the efficiency including hadronic 
interactions is about 10\% lower than without them.

\ack
We would like to thank Dr. S. Banerjee for helpful discussions and 
suggestions. We also acknowledge the contribution of many other past 
members of the BES Collaboration. This work is supported in part by 
the National Natural Science Foundation of China under contracts Nos. 
10491300, 10225524, 10225525, the Chinese Academy of Sciences under 
contract No. KJ 95T-03, the 100 Talents Program of CAS under Contract 
Nos. U-11, U-24, U-25, and the Knowledge Innovation Project of CAS 
under Contract Nos. U-602, U-34 (IHEP); by the National Natural 
Science Foundation of China under Contract No. 10175060 (USTC), and 
No. 10225522 (Tsinghua University); and by the Department of Energy 
under Contract No. DE-FG02-04ER41291 (U Hawaii).


\begin{thebibliography}{00}
\bibitem{mark3} The generators were originally developed by the 
MARKIII Collaboration.
\end{thebibliography}
\end{document}